\documentclass[prd,aps,preprint,amsmath,nofootinbib,amssymb,eqsecnum,showkeys,tightenlines]{revtex4-1}
\usepackage{hyperref} 
\usepackage{slashed}
\usepackage{epsfig,latexsym,cancel,amssymb,amsmath,verbatim,mathrsfs}
\usepackage{color}
\usepackage{graphicx}

\usepackage[inline]{enumitem}
\usepackage[multidot]{grffile}  
\usepackage{dcolumn}
\usepackage{bm}
\usepackage{amsmath}
\usepackage{amsfonts}
\usepackage{amssymb}
\usepackage{color}
\usepackage{latexsym}
\usepackage{slashed} 
\usepackage{pstricks}
\usepackage{indentfirst}
\usepackage{mathrsfs}
\usepackage{multirow}
\usepackage{epsfig,psfrag}
\usepackage{subfigure}
\usepackage{setspace} 
\usepackage[utf8]{inputenc} 
\usepackage[scientific-notation=true]{siunitx} 

\newcommand{\SUtwoL}{\mathrm{SU}(2)_\mathrm{L}}

\allowdisplaybreaks 

\graphicspath{{fig/}}

\begin{document}

\title{  {Minimal asymptotically safe dark matter} }
\author{Chengfeng Cai$^1$}
\author{Hong-Hao Zhang$^{1,}$}\email[]{zhh98@mail.sysu.edu.cn}
\affiliation{$^1$School of Physics, Sun Yat-Sen University, Guangzhou 510275, China}

\begin{abstract}
We study a simple class of dark matter models with $N_f$ copies of electroweak fermionic multiplets, stabilized by $\mathrm{O}(N_F)$ global symmetry. Unlike conventional minimal dark matter which usually suffers from Landau poles, in these models the gauge coupling $g_2$ has a non-trivial ultraviolet fixed point, and thus is asymptotically safe as long as $N_F$ is large enough.
These fermionic $n$-plet models have only two free parameters: $N_F$ and a common mass $M_{\mathrm{DM}}$, which relate to dark matter relic abundance. We find that the mass of triplet fermionic dark matter with $N_F\sim \mathcal{O}(10)$ flavors can be several hundred GeV, which is testable on LHC. A benefit of large $N_F$ is that DM pair annihilation rate in dwarf galaxies is effectively suppressed by $1/N_F$, and thus they can evade the constraint from $\gamma$-ray continuous spectrum observation.
For the case of triplets, we find that the models in the range $3\leq N_F\leq20$ are  consistent with all current experiments.
However, for $N_F$ quintuplets, even with large $N_F$ they are still disfavored by the $\gamma$-ray continuous spectrum.
\end{abstract}

\maketitle

\tableofcontents

\clearpage

\section{Introduction}\label{sect1}
Dark matter (DM) is an astronomical and cosmological indication for new physics beyond the standard model.
An economical way for describing DM is to consider the so-called minimal dark matter (MDM) model~\cite{Cirelli:2005uq,Cirelli:2007xd,Cirelli:2008id,Cirelli:2009uv,Hambye:2009pw,Buckley:2009kv,Cai:2012kt,Earl:2013jsa,Cirelli:2014dsa,Ostdiek:2015aga,Cirelli:2015bda,Garcia-Cely:2015dda,Cai:2015kpa,DelNobile:2015bqo}. In this model, the neutral component of an additional electroweak $\SUtwoL$ multiplet plays the role of DM, which belong to weakly interacting massive particles (WIMPs) and were thermally produced in the early universe~\cite{Bertone:2004pz,Feng:2010gw,Young:2016ala,Arcadi:2017kky}.
The free parameter for this model is only a common DM mass $M_{\mathrm{DM}}$. The mass can be determined by the cosmological relic abundance of DM if we assume this candidate as the whole relic of DM, and thus the model actually has no free parameter.

Recently, there are several experiments hunting the signal from this kind of WIMPs. One is the direct detection which searches for a recoiled signal from DM-nucleon scattering. As the rapid development of the direct detection sensitivity, the current bound on the cross section of spin independent (SI) scattering has reached $\mathcal{O}$($10^{-45}$)~cm$^2$ for mass of about $\mathcal{O}$(TeV)~\cite{Cui:2017nnn,Akerib:2016vxi,Aprile:2018dbl}. The experimental result implies that only the low dimensional representation multiplets ($n\leq5$) are allowed~\cite{Hisano:2015rsa}. Another important experiment is the detection of the continuous spectrum of $\gamma$-ray produced from DM pair annihilation in dwarf galaxies, which has been observed by Fermi-Lat~\cite{Ahnen:2016qkx}. As discussed in Ref.~\cite{Cirelli:2007xd,Chun:2015mka,Mitridate:2017izz}, if the Sommerfeld enhancement effect is included, the MDM models with $M_{\mathrm{DM}}\sim$ $\mathcal{O}$(TeV) have tension with the current bound on the annihilation cross section.
Apart from searching DM signal from the sky, people also try to produce DM particles in colliders. A	promising way to hunt electroweak (EW) multiplets with almost degenerate masses in Large Hadron Collider (LHC) is to detect the disappearing track of long-lived charged particle, which finally decays to a neutral particle and a pion~\cite{Ibe:2006de,Buckley:2009kv}.

In this paper, we will extend the MDM model by introducing $N_F$ EW fermionic multiplets with a common Majorana mass. It means that we only focus on a model of $n=2k+1,~(k=1,2,...)$ dimensional representations of $\textrm{SU}(2)_L$ with hypercharge $Y=0$. In the usual setup of MDM, for the triplet we need to impose a $Z_2$ parity for stablizing the neutral component. In our setup, the dark sector will be protected by an $\textrm{O}(N_F)$ global symmetry in order to prohibit DM decay through renormalizable  and dimension 5 operators. Since this symmetry also reduces the number of parameters, the model is still predictive. Comparing to the MDM  model, we introduce an extra parameter, the flavor number of fermionic fields $N_F$.

When both the flavor number $N_F$ and the dimension of representation $n$ are small, the 1-loop beta-function of $\textrm{SU}(2)_L$ gauge coupling $g_2$ is slightly modified and remains the perturbativity up to the Planck scale. On the other hand, when $N_F$ is large enough, the running of $g_2$ is in the situation called asymptotic safety, which means $g_2$ has an interacting UV fixed point~\cite{Wilson:1971bg,Weinberg:1980gg,Litim:2014uca,Litim:2015iea,Esbensen:2015cjw,Pelaggi:2017abg}.\footnote{Some recent lattice simulations show an early hint of Landau pole for $g_2$ \cite{Leino:2019qwk}, however this study can not exclude the existence of UV fixed point with strong renormalized coupling yet. Further studies on this topic are needed.} As pointed out by Ref.\cite{Pelaggi:2017abg}, although $g_2$ is safe with large number of fermionic multiplets, the Higgs quartic coupling $\lambda$ will suffer a problem of vacuum instability. It is due to the all order of contribution to $g_2^4$ term in $\beta_\lambda$ turns its coefficient into a large negative value. This term makes $\lambda$ drop down to the negative value and induce an unstable vacuum. However, we know that Higgs boson is a special particle in the Standard Model (SM), which leads to many famous problems, e.g., the Hierarchy problem. One of the Beyond Standard Model (BSM) extension for solving this problem is to regard the Higgs boson as a composite state, which is an pseudo Nambu-Goldstone boson (pNGB) arised from some symmetry breaking due to strong dynamics~\cite{Kaplan:1983fs,Kaplan:1983sm}. Therefore the cut-off scale of the running should be TeV scale before $\lambda$ becomes negative.

Introducing $N_F$ duplicates of EW multiplet is also interesting in phenomenology. Since the number of DM candidates in these models is $N_F$ times of a single flavor, their mass should be scaled by a factor $1/\sqrt{N_F}$ with respect to the prediction of MDM model for matching the observation of DM relic abundance. If $N_F$ is large enough, their mass can lie in the reachable energy region of LHC. Extending to $N_F$ flavors also enhances the production rate of the multiplets. Therefore we may expect a stringent bound for our models. We will discuss the constraint on $N_F$ triplets model by the current searches of disappearing track in LHC. Finally we consider the impact of introducing $N_F$ flavors on the DM indirect detection. We find that for the triplet models, it can survive from the stringent constraint from the current observation of the continuous spectrum of $\gamma$-ray~\cite{Ahnen:2016qkx}.

This paper is organized as follows. In Sec.~\ref{sec:model}, we introduce the basic setup of $N_F$ DM model and the running behavior of $g_2$. In Sec.~\ref{sec:phenom}, we will discuss some DM phenomenologies, including the direct detection, the relic abundance, the disappearing track, and the continuous spectrum of $\gamma$-ray. Conclusions are given in Sec.~\ref{sec:concl}.

\section{The Model}\label{sec:model}

The basic setup of $N_F$ Dark Matter Model is introducing $N_F$ copies of $\textrm{SU}(2)_L$ fermionic n-plet with a common mass $M_{DM}$. For simplicity, we only consider the cases that all n-plets are left-handed Weyl spinors with $n=2k+1,~k=1,2,...$ (odd dimension of representation) with hypercharge $Y=0$. The Lagrangian for these n-plet $\Psi^I_i$ (i=1,2,...,n;~I=1,2,...,$N_F$) is as follow
\begin{eqnarray}
\mathcal{L}_{dark}=(\Psi^I_i)^\dag i\overline{\sigma}^\mu D_\mu \Psi^I_i-\frac{1}{2}M_{DM}\Psi^I_i\Psi^I_i+h.c., \label{eq-1}
\end{eqnarray}
where we have used the summation convention for both $I$ and $i$ indices. The covariant derivative term of Eq.~\eqref{eq-1} has an accidental global $\textrm{U}(N_F)$ symmetry, which is broken to a subgroup by the mass terms. For simplicity we assume these fermions have a degenerate Majorana-type mass. That is, a global $\textrm{O}(N_F)$ flavor symmetry is assumed in the dark sector. This flavor symmetry not only reduces the number of parameters, but also forbids the $L\Psi H$ coupling which leads to DM decay in the triplet ($n=3$) case. Therefore the neutral components are stable. Note that since there is only a single interaction, $\textrm{SU}(2)_L$ gauge coupling $g_2$, we can always choose a basis where different flavors do not couple each other. This property is important for the latter discussion on the relic abundance and the indirect detection in Sec.~\ref{sec:phenom}.

When $N_F$ is large enough, the leading term of  $\alpha_2$ beta function is order $N_F\alpha_2^2$ which is a large positive number. It drives $\alpha_2$ growing up rapidly as the energy scale increases, and finally leads to non-perturbative coupling. However, once the next order in $1/N_F$ with all loops corrections have been included, the growing will stop at a UV fixed point. To be precise, the running behavior of $\alpha_2$ can be analyzed by using the formula of beta function for large $N_F$ \cite{Holdom:2010qs,Mann:2017wzh,Pelaggi:2017abg},
\begin{eqnarray}\label{NFbetafunc}
\beta_{\alpha_{2}}&\approx&\frac{\alpha_{2}^{2}}{2 \pi}(-\frac{19}{6}+\Delta b_{2})+\frac{\alpha_{2}^{2}}{3 \pi} F_2\left(\Delta b_{2} \frac{\alpha_{2}}{4 \pi}\right),\\
\Delta b_{2}&=&\frac{2}{3}T(R_f)N_F,\quad F_2(A)\equiv \int_{0}^{A} I_{1}(x) I_{2}(x) dx,\nonumber\\
I_{1}(x) &\equiv& \frac{(1+x)(2 x-1)^{2}(2 x-3)^{2} \sin ^{3}(\pi x) \Gamma(x-1)^{2} \Gamma(-2 x)}{\pi^{3}(x-2)},\nonumber\\
I_{2}(x) &\equiv&\frac{3}{4}+\frac{\left(20-43 x+32 x^{2}-14 x^{3}+4 x^{4}\right)}{(2 x-1)(2 x-3)\left(1-x^{2}\right)},\nonumber
\end{eqnarray}
where $T(R_f)$ is the index for fermions' $\textrm{SU}(2)_L$ representation. It is normalized to $1/2$ for doublet (fundamental), $T(R_f)=2$ for triplet (adjoint), and $T(R_f)=10$ for quintuplet.
As we see from formula \ref{NFbetafunc}, $F_2(A)$ diverges to $-\infty$ when $A\to1$. This means that at some point $A^\ast$ that is very closed to $1$, its contribution cancels the large positive leading term and leads to a UV fixed point, $\alpha_2^\ast\approx 4\pi/\Delta b_2$. For triplets (n=3) model, which is the adjoint representation of SU$(2)_L$, $N_F>7$ is large enough for using these formula according to the conclusion from ref.\cite{Antipin:2017ebo}. Note that if we introduce $N_F$ multiplets at TeV scale, we should make sure that the value $\alpha_2(\textrm{TeV})\approx \frac{1}{30}\lesssim \alpha_2^\ast$, otherwise the resummation is no longer reliable. This condition sets an upper bound on the number $N_F$. For the triplet, it is $N_F\lesssim280$, while for quintuplet, it is $N_F\lesssim56$.  In this work, we will focus on some cases within $9\leq N_F\leq 36$ for triplets since the $N_F\sim \mathcal{O}(10)$ is most interesting in phenomenology. For larger $N_F\sim \mathcal{O}(100)$ would be less attractive since they predict a DM mass as small as $100$~GeV which suffered strong constraints from the current LHC experiments~\cite{Aaboud:2017mpt,Sirunyan:2018ldc}. For quintuplets, we consider several cases in the range $16\leq N_F\leq49$ which are in a range below the upper bound.

\begin{figure}[!t]
	\centering\includegraphics[width=.47\textwidth]{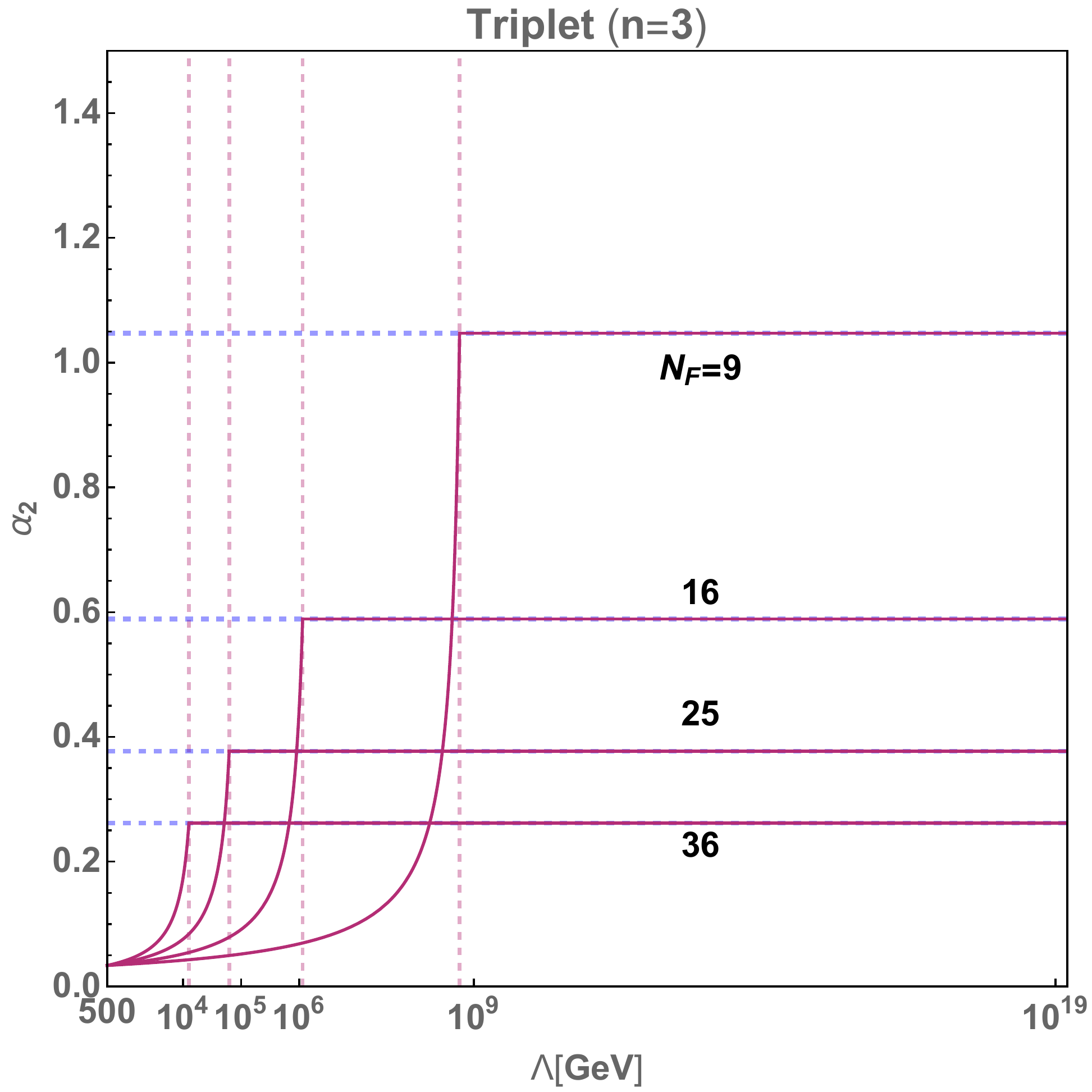}
	\centering\includegraphics[width=.47\textwidth]{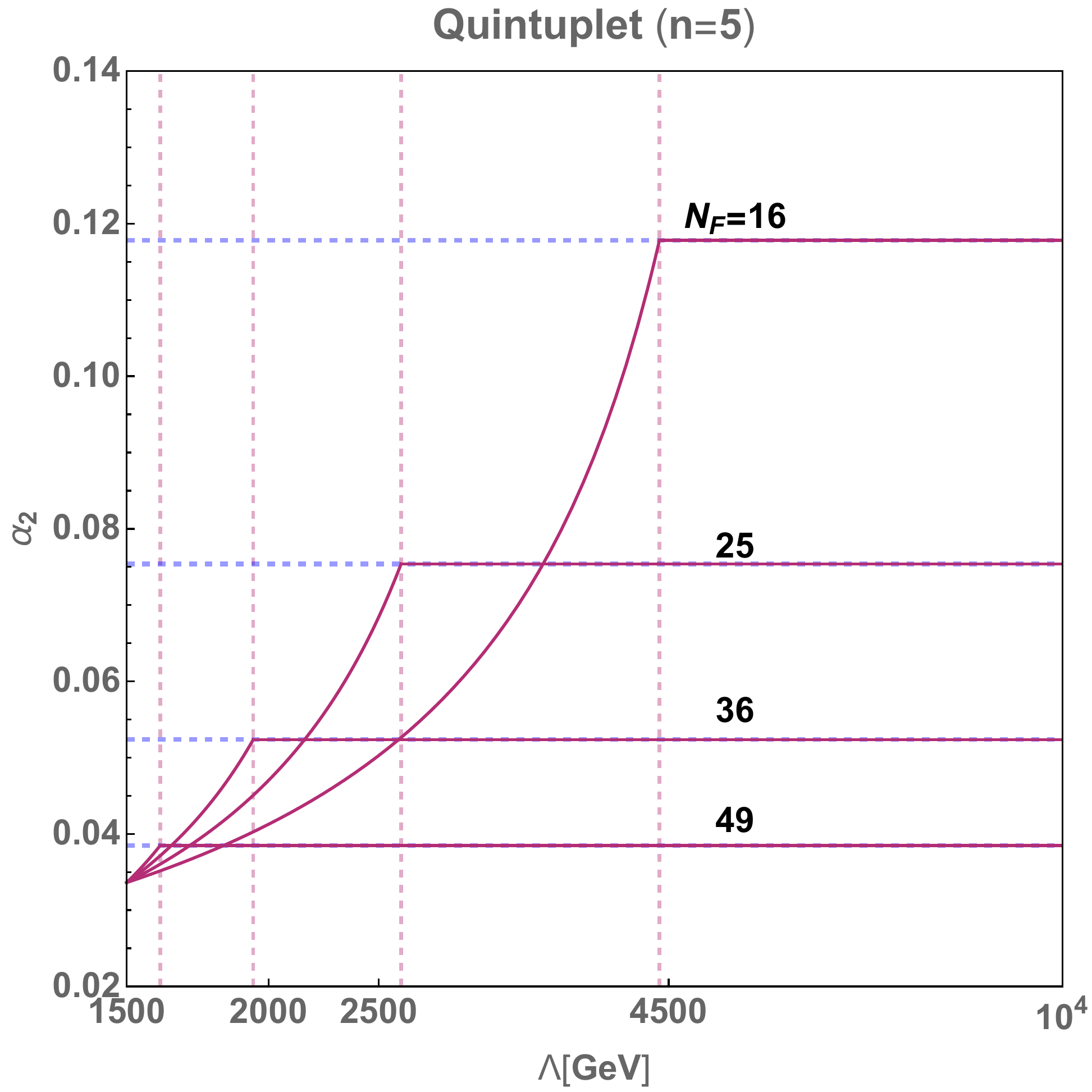}
	\caption{Running of $\alpha_2$ in triplets (left) and quintuplets (right) model with different $N_F$.}\label{asym_safe}
\end{figure}

Note that for triplets, although the cases of $2<N_F\lesssim5$ lose both asymptotic freedom and safety, the gauge coupling $\alpha_2$ still keep perturbativity up to the Planck scale. Thus we can still treat them as an effective theory cutoff at Planck scale. However, for quintuplet, the case of $N_F=1$ loses asymptotic freedom and the cases of $N_F>1$ become non-perturbative in high-energy if $N_F$ is not large enough to acquire asymptotic safety.

A problem raised by these $N_F$ flavors models is that the vacuum becomes unstable in high energy below the Planck scale. In ref.\cite{Pelaggi:2017abg}, the resummed beta function of Higgs quartic coupling $\lambda$ is shown. The authors found that the coefficient of $g_2^4$ term turns to a large negative value at the $g_2$ UV fixed point, thus it will drive the $\lambda$ diving below zero rapidly. They also pointed out that if the fixed point value $\alpha_2^\ast$ is closed to the SM value (needed $N_F\sim 200$ for triplets and $\sim 40$ for quintuplets), the vacuum can be rescued by introducing another scalar field. However for our triplet models with $N_F\lesssim 40$, it seems unlikely to work without introducing other complicated setups. Instead of finding a way to solve this problem, we regard this vacuum instability as a hint that the SM Higgs is just an effective model below TeV scale. This is not a new point of view since people have developed many models to modify the SM Higgs for solving other problems. The most famous example is the hierarchy problem. A minimal extension which can solve the vacuum instability without changing our results might be the $\textrm{SU}(4)/\textrm{Sp}(4)$ composite Higgs model~\cite{Katz:2005au,Gripaios:2009pe,Galloway:2010bp,Barnard:2013zea,Ferretti:2013kya,Cacciapaglia:2014uja,Agugliaro:2016clv,Sannino:2016sfx}. In this model, a new strong dynamics corresponding to a non-abelian group $G_{TC}$ is introduced (the minimal choice is $\textrm{SU}(2)$). Several fermions with this symmetry and the SM gauge symmetries are introduced. They condensate below $10$~TeV and trigger the electro-weak symmetry breaking by vacuum misalignment. Since there is a $\textrm{SU}(2)_L$ doublet techni-fermion in this model, it will contribute a correction to $b_2$: $\Delta b_2=2/3$. In this framework, the Higgs boson is a composite pNGB boson which is invalid in the energy scale over its decay constant, $f\sim1$~TeV. Therefore, the physics of the SM Higgs boson has the cut-off around $1$~TeV and there is no problem with the vacuum stability. \footnote{Note that for the composite Higgs model to obtain the realistic flavor structure, we usually need to introduce some new scalar fields (see Ref.\cite{Agugliaro:2016clv,Sannino:2016sfx} as examples). The quartic coupling of these scalar fields might bring the problem of unstable vacuum back again. However, since the exact value of the quartic couplings are still unknown, it is possible that its value is large enough to cancel the contribution from the $g_2^4$ terms in the beta function.}

In FIG.\ref{asym_safe}, we show the running of $\alpha_2$ with large $N_F$ for triplets and quintuplets as benchmark models. In both cases, we use the SM beta function below the masses threshold $500$~GeV for triplets and $1.5$~TeV for quintuplets. Above the thresholds, we use the resummed beta function defined in eq.\eqref{NFbetafunc}. We find that for triplets with $N_F=9,16,25,36$, $\alpha_2$ reaches its fixed point around $10^{9},10^{6},10^{5},10^{4}$~GeV respectively.  For $N_F=16,25,36,49$ quintuplets, all their $\alpha_2$ fixed points are reached below $10$~TeV.

\section{DM phenomenology}\label{sec:phenom}
In this section, we discuss DM phenomenology. Especially, we will only focus on the triplets (n=3) and quintuplets (n=5) cases because only triplets and quintuplets still survive from the current experimental constraints by the direct detection. The processes leading to the DM-nucleon scattering for triplets and quintuplets are generated by 1-loop gauge mediation, as in the case with the MDM model. The scattering cross section with QCD effects for $n=2,3,5$ have been discussed in ref.\cite{Hisano:2015rsa}. They showed that the theoretical cross section is $\mathcal{O}(10^{-47})$~cm$^2$ for $n=3$ and $\mathcal{O}(10^{-46})$~cm$^2$ for $n=5$. In the interested mass region, triplet's cross section is smaller than the bound from current direct detection experiments (PANDAX, LUX and XENON1T)~\cite{Cui:2017nnn,Akerib:2016vxi,Aprile:2018dbl} by an order of magnitude, while the case of the quintuplet shows just below the bound. We can expect that $n>5$ models with $\mathcal{O}(\textrm{TeV})$ masses have been excluded by direct detection.

As we have mentioned, different flavors do not co-annihilate in these models. Therefore the relic abundance of DM is simply $N_F$ times of that predicted by a single flavor. As we known, the leading order of the effective annihilation cross section $\langle\sigma v\rangle$ for a single n-plet is approximated as~\cite{Cirelli:2005uq,Mitridate:2017izz}:
\begin{eqnarray}
\langle\sigma v\rangle_{eff}\approx\frac{g_{2}^{4}\left(2 n^{4}+17 n^{2}-19\right)}{1536\pi  M_{DM}^{2}}.
\end{eqnarray}
From this formula, we can see that the cross section is proportional to $1/M_{DM}^{2}$. The total relic abundance of DM can be estimated by
\begin{eqnarray}
\Omega_{\mathrm{DM}} h^{2} \approx N_F\times\frac{1.07 \times 10^{9} \mathrm{GeV}^{-1}}{\sqrt{g_{*}} M_{\mathrm{Pl}}\int_{x_{\mathrm{F}}}^\infty\frac{\langle\sigma v\rangle_{eff}}{x^2}dx}\propto N_FM_{DM}^2.
\end{eqnarray}
In order to match the current observation of relic abundance, $M_{DM}$ should be smaller by factor $1/\sqrt{N_F}$ in comparison with a single n-plet. Moreover the Sommerfeld enhancement (SE) effect should be included~\cite{Cirelli:2007xd} to get a more accurate result.
Without the SE effect, the masses for MDM implied by relic abundance are $2.4$~TeV and $4$~TeV for triplets and quintuplets respectively. If this effect is considered, triplet mass increases to about $2.9$~TeV, while quintuplet mass increases to about $9$~TeV. Furthermore, a more complete analysis including bound-state effects~\cite{Mitridate:2017izz} provides a significant correction for quintuplet and the DM mass increases to $11$~TeV. In this work, we will not consider DM pair forming bound state since the quintuplet mass region we are interested in is below $3$~TeV, where the bound state formation effect is not significant comparing to SE.
\begin{figure}[!h]
	\centering\includegraphics[width=.47\textwidth]{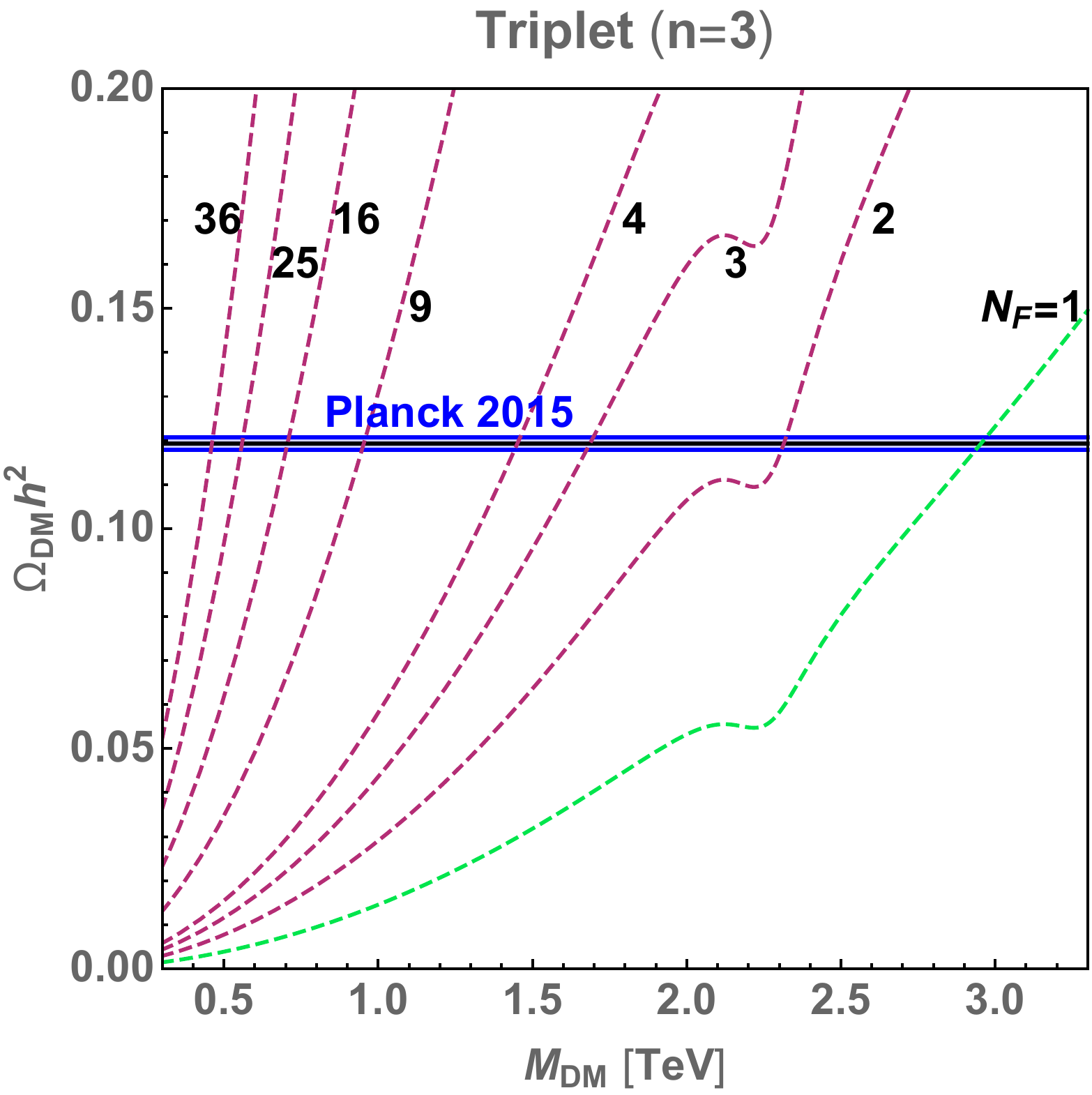}
	\centering\includegraphics[width=.47\textwidth]{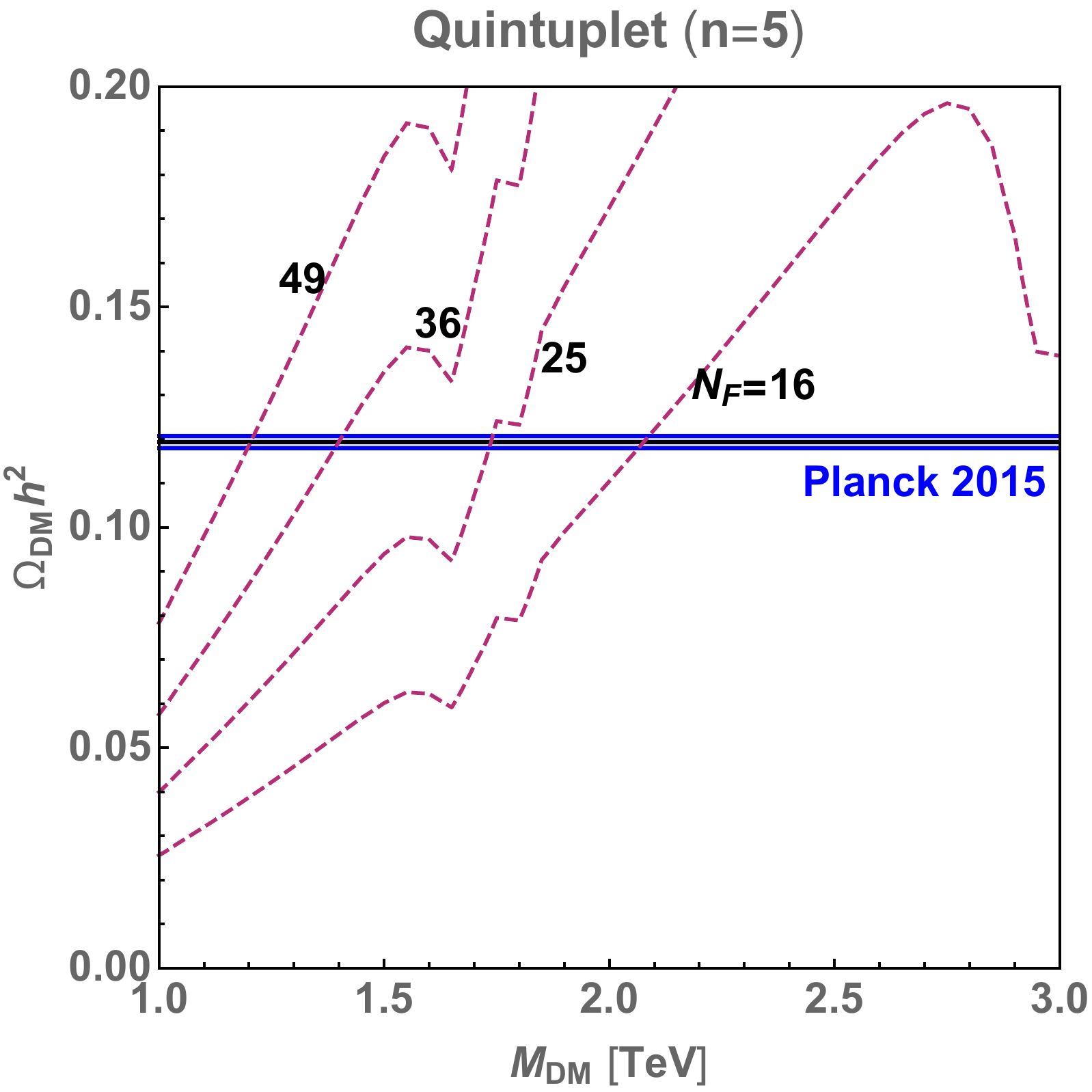}
	\caption{DM relic abundance of triplets (left) and quintuplets (right) models. The magenta dashed curves are given by scaling the single flavor prediction with a factor $N_F$. The observed constraint $\Omega_{DM} h^2=0.1193\pm0.0014$ from PLANCK 2015~\cite{Ade:2015xua} is shown as blue band.}\label{relic_ab}
\end{figure}

In FIG.\ref{relic_ab}, we show the predicted relic abundance of EW triplets and quintuplets models with different $N_F$. The observation results $\Omega_{DM}h^2=0.1193\pm0.0014$ (TT,TE,EE + LowP + lensing) from the PLANCK experiment~\cite{Ade:2015xua} is also shown in both panels. We can see that for triplets, $N_F\geq9$ model has DM mass smaller than $1$~TeV, while $N_F=2,3,4$ models have DM masses larger than $1$~TeV. For quintuplets model, DM mass is always larger than $1$~TeV for $N_F<50$.

As the masses of DM have been fixed by the relic abundance for each $N_F$, in the following discussion we relate $N_F$ to the DM mass by the relation $N_F=\Omega_{tot}/\Omega_I$, where $\Omega_{tot}$ is the total relic abundance from the observation and $\Omega_I$ is the abundance of a single flavor.

\begin{figure}[!h]
	\centering\includegraphics[width=.47\textwidth]{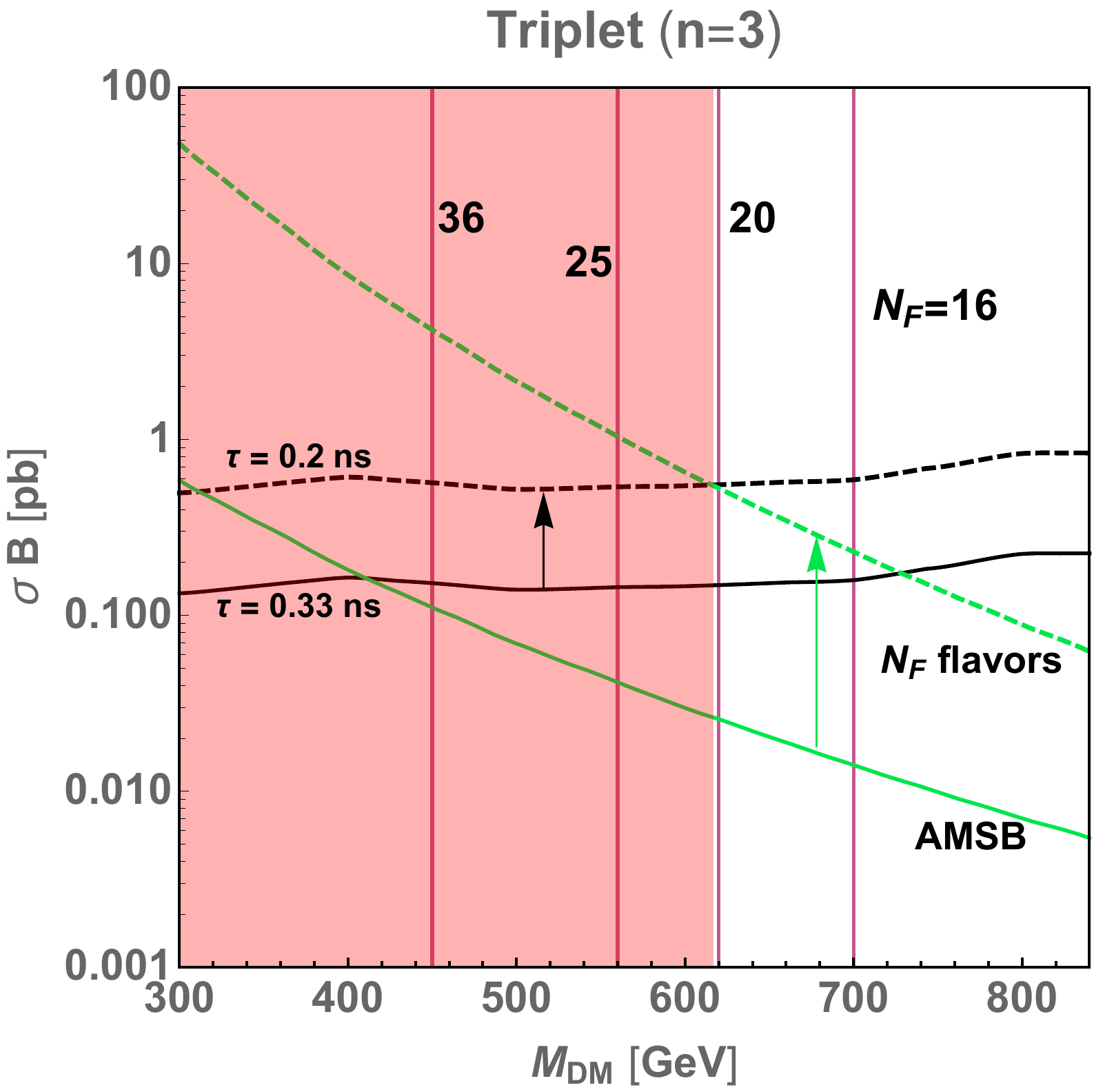}
	\caption{Constraint from disappearing track searches for triplets model. The vertical magenta lines are the masses fixed by the relic abundance for different $N_F$. The black dashed curve is the observed limit of cross section corresponding to $0.2$~ns as the lifetime of charged state. The green dashed curve is the theoretical prediction of $N_F$ triplets models. The red shaded region is excluded with 95\% C.L. by CMS based on integrated luminosity of $38.4$~fb$^{-1}$ at $13$~TeV pp-collision.~\cite{Sirunyan:2018ldc}.}\label{disap_track}
\end{figure}

Next we consider the constraint on triplets model from collider searching for the disappearing track of the long-living charged particle. Both ATLAS and CMS collaborations have studied such signal~\cite{Aaboud:2017mpt,Sirunyan:2018ldc} by assuming that some charginos are produced in the proton-proton collision and traveling some distance before decaying to neutralinos and pions. In the $N_F$ triplets model, every flavors have the same behavior as the wino. The lifetime of charged state is about $0.2$~ns~\cite{Feng:1999fu} which corresponds to the mass splitting about $0.164$~GeV when $M_{DM}$ is larger than $300$~GeV~\cite{Ibe:2012sx}. In the searches from ATLAS based on integrated luminosity of $36.1$~fb$^{-1}$ at $13$~TeV, they set a lower bound on chargino's mass about $460$~GeV~\cite{Aaboud:2017mpt} with $95\%$ C.L. A similar search is also done by CMS based on integrated luminosity of $38.4$~fb$^{-1}$, and the $95\%$ C.L. exclusion bound from their study is about $310$~GeV~\cite{Sirunyan:2018ldc} for lifetime $0.2$~ns. In both groups' studies, electroweak production of charged states are considered. Since we have already assumed in our models, we can use their results to put a constraint on our $N_F$ triplets model. In the first panel of Figure 3 from ref.\cite{Sirunyan:2018ldc}, they show a plot of the observed limit of the cross section and the prediction of AMSB model for chargino with lifetime $0.33$~ns as a benchmark. In order to estimate the observed limit of cross section for the lifetime $0.2$~ns (which is not presented in \cite{Sirunyan:2018ldc}), we scale their observed limit curve with a factor $3.7$ to cross the theoretical line at the mass $\sim310$~GeV (the mass bound given by their Figure 4). In FIG.\ref{disap_track}, we show the observed limit curve for lifetime $0.33$~ns (black solid curve) and the estimated observed limit for lifetime $0.2$~ns (black dashed curve). We also extract the theoretical prediction from ref.\cite{Sirunyan:2018ldc} and scale the curve with a mass dependent factor $N_F(M_{DM})=\frac{\Omega_{tot}^{(obs.)}}{\Omega_I(M_{DM})}$ to obtain our prediction for $N_F$ triplets model. The predicted curves from the AMSB (green solid curve) and $N_F$ flavors (green dashed curve) are shown in the same figure. The mass bound for the $N_F$ triplets model can be found at the crossing point between the black dashed curve and the green dashed curve. It shows that triplet models with $N_F>20$ are excluded, which corresponds to a DM mass lower bound $\sim620$~GeV.

Finally, we discuss the constraint from the current observation of the $\gamma$-ray continuous spectrum~\cite{Ahnen:2016qkx}. A benefit of considering $N_F$ flavors in our model is to loose the tension between the theoretical prediction and the observation. There are two reasons. One is that, in our model with a proper $N_F$, the DM mass will be scaled away from the resonance of Sommerfeld enhancement effect. The second reason is that, for each flavor of DM, their number density in the dwarf galaxies is only $1/N_F$ times as much as the total DM number density. Since there is no co-annihilation between different flavors, the probability of the annihilation of a pair of DM particles is suppressed by a factor of $1/N_F^2$. Therefore the total $\gamma$-ray flux from $N_F$ flavors will be suppressed by a factor $1/N_F$. We will see that such suppression can make the triplets model with $N_F>2$ consistent with the current observation.

\begin{figure}[!h]
	\centering\includegraphics[width=.47\textwidth]{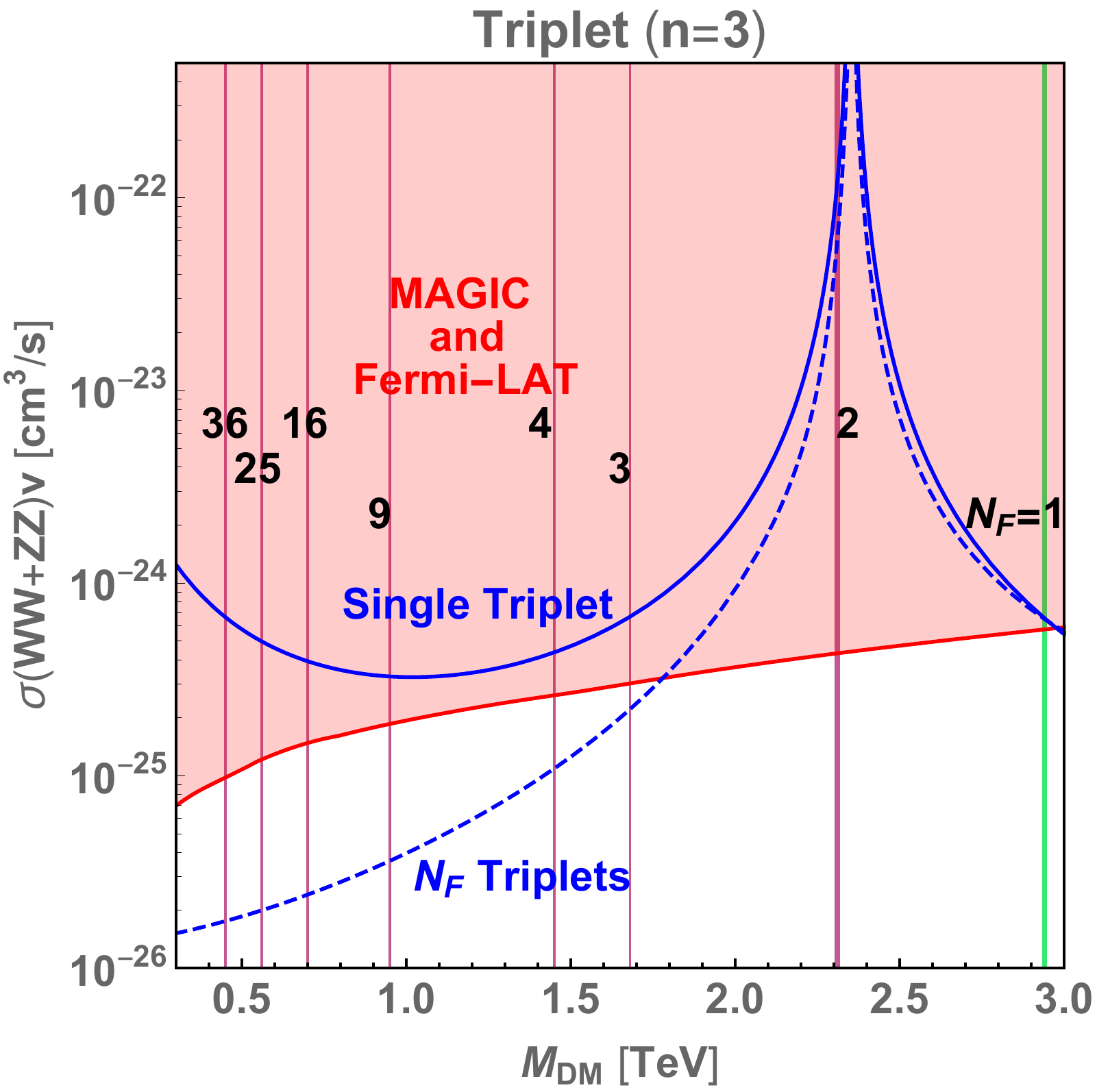}
	\centering\includegraphics[width=.47\textwidth]{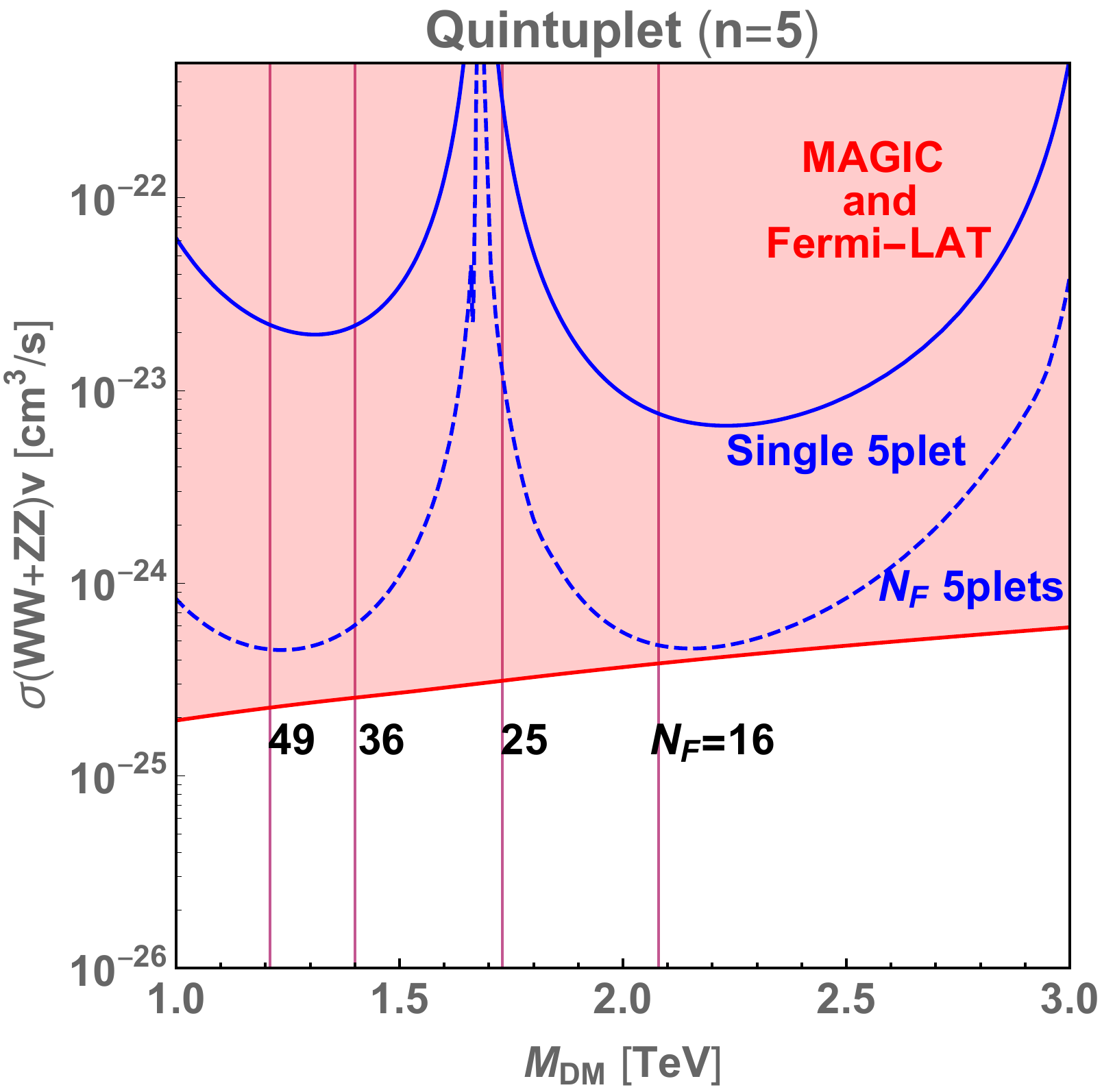}
	\caption{Constraints from $\gamma$-ray spectrum observation for triplets (left) and quintuplets (right). The vertical magenta lines are the masses fixed by the relic abundance for different $N_F$. The blue dashed curves are the prediction of $N_F$ flavors. The red shaded regions are excluded at 95\% C.L.~\cite{Ahnen:2016qkx}}\label{ind_det}
\end{figure}

The observed $\gamma$-ray flux produced by $N_F$ DM annihilation in a given region of sky ($\Delta\Omega$) can be evaluated by~\cite{Ahnen:2016qkx}
\begin{eqnarray}\label{flux1}
\frac{d \Phi}{d E}(\Delta \Omega)=\frac{1}{4 \pi}\sum_{I=1}^{N_F} \frac{\langle\sigma v\rangle_I J_I(\Delta \Omega)}{2 M_{\mathrm{DM}}^{2}} \frac{d N_I}{d E},
\end{eqnarray}
where $\frac{d N_I}{d E}$ is the averaged $\gamma$-ray spectrum per annihilation event of the $I$-th flavor. The J-factor for each flavor is defined as
\begin{eqnarray}
J_I(\Delta \Omega)=\int_{\Delta \Omega} d \Omega^{\prime} \int_{\mathrm{l.o.s}} d l \rho_I^{2}\left(l, \Omega^{\prime}\right),
\end{eqnarray}
where $\rho_I(l,\Omega')$ is the $I$-th flavor DM mass density. Since all flavors have a common mass and no co-annihilation between different flavors in our model, each $\rho_I$ is just $1/N_F=\Omega_I/\Omega_{tot}$ of the total density $\rho_{tot}$. The cross section and spectrum per annihilation for all flavors are also the same as a single flavor model. When we use \eqref{flux1} to compare with the observation, it is convenient to write it as
\begin{eqnarray}
\frac{d \Phi}{d E}&=&N_F\times \frac{1}{N_F^2}\frac{1}{4 \pi} \frac{\langle\sigma v\rangle J_{tot}(\Delta \Omega)}{2 M_{\mathrm{DM}}^{2}} \frac{d N}{d E}=\frac{1}{4 \pi} \frac{\left(\frac{1}{N_F}\langle\sigma v\rangle\right) J_{tot}(\Delta \Omega)}{2 M_{\mathrm{DM}}^{2}} \frac{d N}{d E},
\end{eqnarray}
 which is effectively equivalent to scale the theoretical annihilation cross section suppressed by a factor of $1/N_F$ when we use this formula to compare with the observation. This suppressing factor is significant when $N_F$ is large. The final states from DM pair annihilation we need to calculate is $W^+W^-$ and $ZZ$. In FIG.\ref{ind_det}, we show the constraints on both the triplets (left panel) and the quintuplets (right panel) models. We show the annihilation cross section predicted by a single flavor (blue solid curve) and $N_F$ flavors (blue dashed curve) for comparison. The blue dashed curves are plotted by scaling the blue solid curves with a mass dependent factor $1/N_F=\Omega_I(M_{DM})/\Omega_{tot}^{(obs.)}$. We can see that both the MDM ($N_F=1$) and $N_F=2$ triplets models are disfavored by the observation due to the SE resonance around 2.4~TeV, while models with $N_F>2$ survive. For the quintuplets with large $N_F$, all interesting regions are still disfavored by the experiment even with the $1/N_F$ suppression.

\section{Conclusions}\label{sec:concl}

In this work, we study a DM model with $N_F$ copies of fermionic EW multiplets. DM stability can be guaranteed by an O($N_F$) global symmetry. These models are quite predictive since we introduce only one free parameter, the flavor number $N_F$. Once the flavor number is fixed, the DM mass can be determined by the observation of the DM relic abundance. Since the whole DM relics are constituted by $N_F$ flavors, each flavor only contributes $1/N_F$ of the total abundance. This allows the DM mass to be much lighter than the value predicted by MDM models if $N_F$ is very large. On the other hand, it is reasonable and motivative to consider a large $N_F$ since they can have the asymptotic safety of the gauge coupling $g_2$.

We focus on triplets and quintuplet in this work for avoiding the stringent bound from the direct detection. We find in the case of triplets DM that the masses could be about $400,~560,~700,$ and $950$~GeV for $N_F=36,~25,~16$, and $9$ respectively. These regions could be tested in the current and future high-luminosity collider experiments.
We examine the $N_F$ triplets model by the LHC searches of the disappearing track of the long-lived particle with $38.4$~fb$^{-1}$ ($13$~TeV) from CMS experiment. We find that triplets with $N_F>20$ ($M_{DM}<620$~GeV) have been excluded.
 For small $N_F$ ($=2,~3,~4$) flavors of triplets whose asymptotic safety are lost, the allowed masses are above $1$~TeV. For the quintuplets, the allowed masses are about $1.2,~1.4,~1.7,~2.1$~TeV corresponding to $N_F=49,~36,~25,~16$ respectively. In both models the mass regions above TeV are not constrained by the current collider experiments.

For indirect detection, we consider the constraints from the observation of $\gamma$-ray continuous spectrum from dwarf galaxies. We find that the DM pair annihilation rate of $N_F$ flavors model is effectively suppressed by $1/N_F$ since each flavor only constitutes $1/N_F$ of the DM density. As a result, triplets with $N_F>2$ survive under the current bound.
Combining all the constraints on triplets model, the available flavor number is in the range $3\leq N_F\leq20$ corresponding to the mass within $620~\textrm{GeV}<M_{DM}\lesssim1.7$~TeV.
For the quintuplets, the constraint from $\gamma$-ray continuous spectrum is so stringent that all models with large $N_F$ are excluded even they have a $1/N_F$ suppression.

\begin{acknowledgments}
We would like to thank Z.~H. Yu and S. Enomoto for helpful discussions. We also thank G. Cacciapaglia and V. Shahram for drawing our attention to the interest in asymptotic safety of gauge couplings. This work is supported by the National Natural Science Foundation of China (NSFC) under Grant No. 11875327, the China Postdoctoral Science Foundation under Grant No. 2018M643282, the Natural Science Foundation of Guangdong Province under Grant No. 2016A030313313, and the Sun Yat-Sen University Science Foundation.
\end{acknowledgments}


\bibliographystyle{utphys}
\bibliography{ref}

\end{document}